\documentclass[12pt]{article}
\usepackage{amsmath,latexsym,epsfig,amssymb}

\def\bea{\begin{eqnarray}}
\def\eea{\end{eqnarray}}

\newcommand{\be}{\begin{equation}}
\newcommand{\ee}{\end{equation}}
\def\ma[#1,#2,#3,#4]  {{\left( \matrix{ #1  & #2 \cr
                                        #3  & #4 \cr } \right)}}

\begin{document}
\title{
{\vspace{-1cm} \normalsize
\hfill \parbox{40mm}{CERN-TH/2000-196}}\\
{\vspace{-0.5cm} \normalsize
\hfill \parbox{40mm}{IFIC-00/41}}\\
{\vspace{-0.5cm} \normalsize
\hfill \parbox{40mm}{FTUV-00-0712}}\\[25mm]
A note on the practical feasibility of domain-wall fermions \footnote{Talk given by 
P. Hern\'andez at 
the workshop on {\em Current theoretical problems
in lattice field theory}, 2 -- 8 April 2000, Ringberg, Germany.}
\addtocounter{footnote}{+2}
\author{
Pilar Hern\'andez\footnote{On leave from Departamento de F\'{\i}sica Te\'orica, Universidad de Valencia, Spain.},
Karl Jansen\footnote{Heisenberg Foundation Fellow.}
$\;$ and Martin L\"uscher\footnote{On leave from Deutsches Elektronen-Synchrotron DESY, D-22603 Hamburg, Germany.} 
\\
CERN, Theory Division, CH-1211 Geneva 23, Switzerland}}
%\date{\today}
\maketitle

\begin{abstract}

  Domain-wall fermions preserve chiral symmetry up to terms that
  decrease exponentially when the lattice size in the fifth dimension is taken
  to infinity. The associated rates of convergence are
  given by the low-lying eigenvalues of a simple local operator
  in four dimensions. These can be computed using the Ritz functional technique
  and it turns out that the convergence tends to be extremely slow
  in the range of lattice spacings relevant to large-volume
  numerical simulations of lattice QCD. Two methods to improve on
  this situation are discussed. 

\end{abstract}

\pagebreak

{\em Introduction}\\

The idea to realize 4-dimensional (4D) chiral fermions on the lattice by coupling 5D fermions to a 4D domain wall \cite{kaplan} has attracted a lot of attention in the lattice community (for a review see \cite{blum0}).
Although originally designed to construct chiral gauge theories, 
the idea can also be applied to lattice QCD in order to preserve the global chiral symmetry at zero quark mass \cite{shamir1,shamir2}. In this case, 
a 5D Wilson--Dirac operator is chosen 
with $N$ slices in the extra dimension and appropriate Dirichlet boundary
conditions in the fifth dimension. In the limit $N \rightarrow \infty$,  
chiral zero modes exist as surface modes on the 4D boundaries, even at
finite lattice spacing. 

 The bulk fermionic degrees of freedom 
are massive and can be shown to decouple in the continuum limit:  the 
action of the 5D system is equivalent to the one corresponding to  
a 
4D Dirac operator describing the boundary chiral modes \cite{neub3}; similarly, the 
propagator of the boundary fields can be obtained from the inverse of the same 4D operator \cite{kiku1}.  The chirality of the surface modes in the limit $N\rightarrow\infty$ then follows \cite{kiku1} from the fact that, in this limit, this 4D Dirac operator satisfies the Ginsparg--Wilson (GW) relation  \cite{gw}--\cite{neub2}, 
which implies an exact lattice chiral symmetry \cite{luscher}.   
Thus the 5D domain-wall construction in the limit $N\rightarrow\infty$ is 
completely equivalent to a 4D lattice formulation of Ginsparg--Wilson fermions, 
and satisfies all the properties that follow from the exact chiral symmetry \cite {hasen2, hasen, hasen3, luscher, chandra}.
Moreover, if the continuum limit is taken in the extra dimension, this 4D formulation coincides with that using Neuberger's fermion operator \cite{neub1,neub2}.

The introduction of an extra dimension makes 
domain-wall fermions more demanding numerically than standard Wilson fermions (the equivalent 4D formulation is similarly more demanding, owing to the non-ultralocality of the action). Nevertheless the advantage of preserving an exact chiral symmetry 
might compensate for the higher cost in some cases. 
An analysis in the free theory
showed that the convergence to the exact operator as a function 
of $N$ is rapid \cite{shamir1,shamir3}. This gave rise to the hope
that also in the interacting case domain-wall fermions could be used
without too much computational overhead. 
However, in realistic simulations,
there are indications that the convergence
rate deteriorates rapidly at large values of the gauge coupling, 
and much larger values of $N$ are indeed needed \cite{blum1}--\cite{cppacs}, leading
to a substantial computational cost. 

In a recent paper \cite{scri}, the problems found in practical simulations
were traced back to the appearance of very small eigenvalues 
of a certain 4D operator, which control
the rate of convergence in $N$. We have performed an independent study 
and confirm 
the analysis in \cite{scri}. In addition, we discuss a new method to  
improve the domain-wall
fermion operator, which differs from the one proposed in \cite{scri} 
and proves to work better numerically. 

\vspace{0.8cm}

{\em Five dimensional theory}

\vspace{0.8cm}

In this section we establish our notation and collect some useful
formulae, the derivation of which can be found in \cite{shamir1,shamir2,neub3,kiku1,borici}.
The 5D domain wall operator we consider here is defined as
\begin{equation}
\mathcal{D} =\frac{1}{2}\left\{\gamma_5(\partial_s^{*}+\partial_s) - a_s\partial_s^{*}\partial_s\right\}
            +M,\;  
\label{5doperator}
\end{equation}
where $s$ denotes a lattice site in the fifth direction ($a_s \le s \le a_s N$), $a_s$ is the corresponding lattice spacing, and 
$\partial_s^{*}$ and $\partial_s$ are the usual forward and
backward derivatives.    
The operator $M$ in eq.~(\ref{5doperator}) is obtained from the standard 4D Wilson operator by
\begin{equation}
M=D_\mathrm{w} - m_0
\label{Moperator}
\end{equation}
with 
\begin{equation}
D_\mathrm{w} =\frac{1}{2}\left\{\gamma_\mu(\nabla_\mu^{*}+\nabla_\mu)-a\nabla_\mu^{*}\nabla_\mu\right\}.\;
\label{Doperator}
\end{equation}
Here $\nabla_\mu^{*}$ and $\nabla_\mu$ are the gauge covariant forward and backward derivatives and $a$ is the lattice spacing in the four physical dimensions 
$\mu=1,...,4$.  The mass parameter $m_0$ obeys
\begin{equation}
m_0 > 0,\;\;\;
a_s m_0 < 2,\;\;\;
a m_0< 2. \; 
\label{m0range}
\end{equation}
Note that the lattice spacings $a_s$ and $a$ can be different. 
The boundary conditions are fixed through
\begin{eqnarray}
P_+ \psi(0,x) = P_- \psi(a_s N + a_s , x) = 0\;,
\end{eqnarray}
where $P_\pm \equiv \frac{1}{2}(1\pm\gamma_5)$.

Appropriate interpolating fields for the quarks constructed out of the left and right boundary modes are  
% the physical quark fields are given through the boundary fields
\begin{eqnarray}
q(x) & = & P_{-}\psi(a_s,x)+P_{+}\psi(N a_s,x) \label{qofx}\; , \\
\bar{q}(x) & = & \bar{\psi}(a_s,x) P_{+} + \bar{\psi}(N a_s,x)P_{-} \label{qbarofx}. 
\end{eqnarray}
 A mass term can then be introduced by adding to eq.~(\ref{5doperator}) the term 
\begin{equation}
\frac{1}{2} m\left\{\bar{\psi}(a_s,x)P_{+}\psi(N a_s,x)+\bar{\psi}(N a_s,x)P_{-}\psi(a_s,x)\right\}\; .
\label{4dmassterm}
\end{equation} 

The two-point function of the quark fields is related to an effective 4D operator $D_N$ by \cite{kiku1}
\begin{equation}
\langle q(x)\bar{q}(x)\rangle = \frac{2-aD_N}{aD_{m,N}},
\label{twopoint}
\end{equation}
with
\begin{equation}
D_{m,N} = (1-\frac{1}{2}am)D_N+m. 
\label{DmN}
\end{equation}
In terms of the operators $K_{\pm}$,
\begin{equation}
K_\pm \equiv \frac{1}{2}\pm\frac{1}{2}\gamma_5\frac{a_s M}{2+a_sM}, \;
\label{Kpm}
\end{equation}
$D_N$ is given by
\begin{equation}
a D_N = 1 + \gamma_5\frac{K_{+}^{N} - K_{-}^{N}}{K_{+}^{N}+K_{-}^{N}}.  
\label{DN}
\end{equation}

From eq.~(\ref{DN}) it is straightforward to show that 
\begin{equation}
aD\equiv\lim_{N\rightarrow\infty} aD_N = 1+\gamma_5\epsilon(K_{+}-K_{-})\; ,
\end{equation}
which can also be written as \cite{borici}
\begin{eqnarray}
aD & = & 1 - A(A^\dagger A)^{-1/2} \label{D} \\
A & = & -a_s M(2+a_sM)^{-1}\; .\label{A} 
\end{eqnarray}

It follows easily from this expression that $D$ satisfies the Ginsparg--Wilson relation,  the only difference
to Neuberger's operator being the different definition of $A$. Indeed, 
Neuberger's operator is readily obtained from eqs. (\ref{D}) and (\ref{A}) by 
taking the limit $a_s \rightarrow 0$.  

Similarly, in the limit $N\rightarrow \infty$, the fermion determinant of the 5D formulation can be written
in terms of the determinant of $D_{m,N}$, up to local subtractions. The 5D
formulation is thus completely equivalent to a 4D lattice formulation of
Ginsparg--Wilson fermions satisfying an exact chiral symmetry. 

A final necessary condition for this formulation to be an acceptable regularization of QCD is that the 
operator of eq.~(\ref{D}) is local. Indeed, it 
has been shown by Kikukawa \cite{kiku2} that both the operators in eqs. (\ref{D}) and 
(\ref{DN}) are exponentially localized for smooth enough gauge fields, satisfying a plaquette bound \cite{hjl,martin}. 

In realistic simulations of domain-wall fermions, $N$ is finite. 
In this situation, the chiral symmetry is broken by the residual terms $\delta D \equiv D_N -D$.   
It may be speculated that one could include a small 
additive quark mass renormalization, in order
to get rid of these chirality breaking effects. This is, however, only justified by universality
arguments if the subleading corrections in $N$ in the action are local. 
The result of \cite{kiku2} shows that this is indeed the case 
since $\delta D$ is also local. However it is important to 
stress that the exponential localization of $\delta D$ only sets in at 
distances of $\mathrm{O}(N)$. This can be shown already in the free case. On the other hand, in practical simulations the typical lattice
sizes used are not much larger than $N$ and consequently $\delta D$ is not 
local at the distances probed.   
In this situation, a quark mass renormalization cannot cancel the chirality-breaking effects 
induced by $\delta D$. 

\vspace{0.8cm}

{\em The convergence rate in $N$}

\vspace{0.8cm}

For gauge field configurations with 
a restricted plaquette value,   
the operator $A^\dagger A$ has a spectral gap \cite{hjl}: 
\begin{equation}
0 < u \leq A^\dagger A \leq v,  
\end{equation}
 ensuring the exponential convergence in $N$ of $D_N$. The minimum rate of convergence is given by
\begin{equation}
\omega = \mathrm{min_i}[\omega_i], \;\;\;\;\; 
\omega_i \equiv \mathrm{ln}\frac{1+\sqrt{\lambda_i}}{|1-\sqrt{\lambda_i}|}, 
\label{rate}
\end{equation}
where $\lambda_i$ are the eigenvalues of $A^\dagger A$. 

However, in realistic simulations the plaquette bound is not satisfied  
and  it is important to study the convergence rate
$\omega$ for the values of $\beta$ and $m_0$ at which  
large-scale numerical simulations are performed nowadays. 

%The fact that the convergence rate is determined by an only 
%4-dimensional eigenvalue problem renders the task of 
%computing $\omega$ numerically relatively easy:
%generating (quenched) gauge field configurations using standard
%techniques, we have studied the 
%the low-lying and largest
%eigenvalues of $A^\dagger A$. 

The eigenvalues $\lambda$, which determine $\omega$, can be obtained through the generalized 4D eigenvalue equation
\begin{equation}
a_s^2 M^\dagger M\psi=\lambda(2+a_sM)^\dagger(2+a_sM)\psi. \;
\label{genritz}
\end{equation}
It is either the minimum or maximum eigenvalue of $A^\dagger A$ that minimizes
$\omega_i$.  These eigenvalues can be obtained 
 by minimizing (maximizing) the generalized 
Ritz functional
\begin{equation}
\frac{\langle \psi | a_s^2 M^\dagger M|\psi\rangle}{\langle \psi |(2+a_sM)^\dagger(2+a_sM)|\psi\rangle}\; 
\label{ritz}
\end{equation}
 using a straightforward generalization of the 
algorithm described in \cite{cg} \footnote{A more detailed description of
the algorithm can be obtained from the authors on request.}. 

Eigenvalues above the lowest one can be computed by modifying the operator $M^\dagger M$ in the numerator of eq. (\ref{ritz}) in such a way that the already  
computed eigenvalues are shifted to larger values. For example, this can be achieved by substituting $M^\dagger M$ by $M^\dagger M + \sum_i \alpha_i M^\dagger M|\psi_i\rangle\langle\psi_i|(2+a_sM)^\dagger(2+a_sM)$ 
with $\alpha_i \equiv (1-\lambda_i)/\lambda_i$ and $\lambda_i$, $\psi_i$ being the already computed eigenvalues and vectors.
Notice that in this method no inversion of the matrix $(2+a_sM)^\dagger(2+a_sM)$ is needed. 

\begin{figure}
\vspace{0.0cm}
\begin{center}
\psfig{file=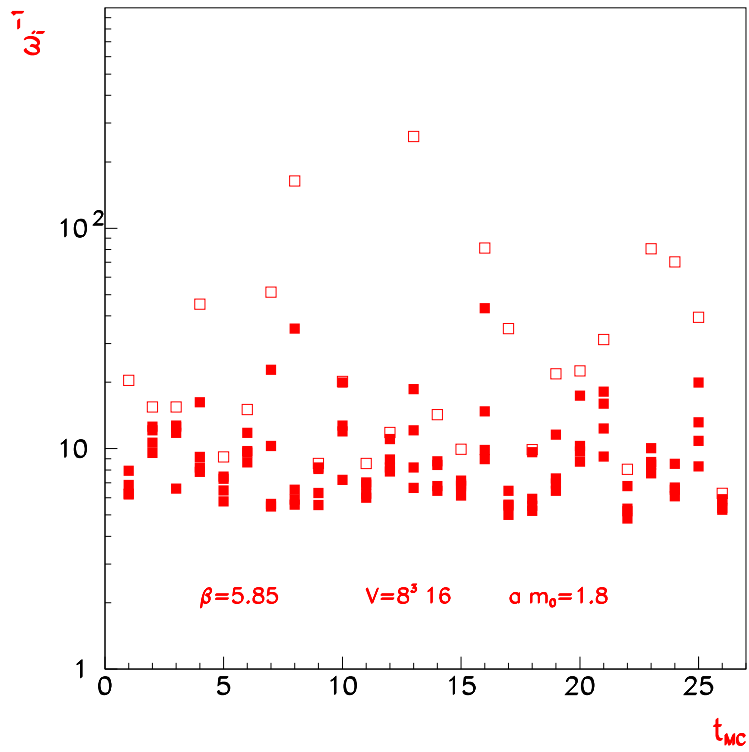, %
width=11cm,height=11cm}
\end{center}
\caption{ \label{fig:eigb585} Inverse convergence rate $\omega_i^{-1}$  for the five lowest eigenvalues of $A^\dagger A$ (open symbols correspond to the 
lowest eigenvalue) as a function of Monte Carlo time $t_{\rm MC}$, at $\beta=5.85$ on an $8^3\cdot 16$ lattice.}
\end{figure}

\begin{figure}
\vspace{0.0cm}
\begin{center}
\psfig{file=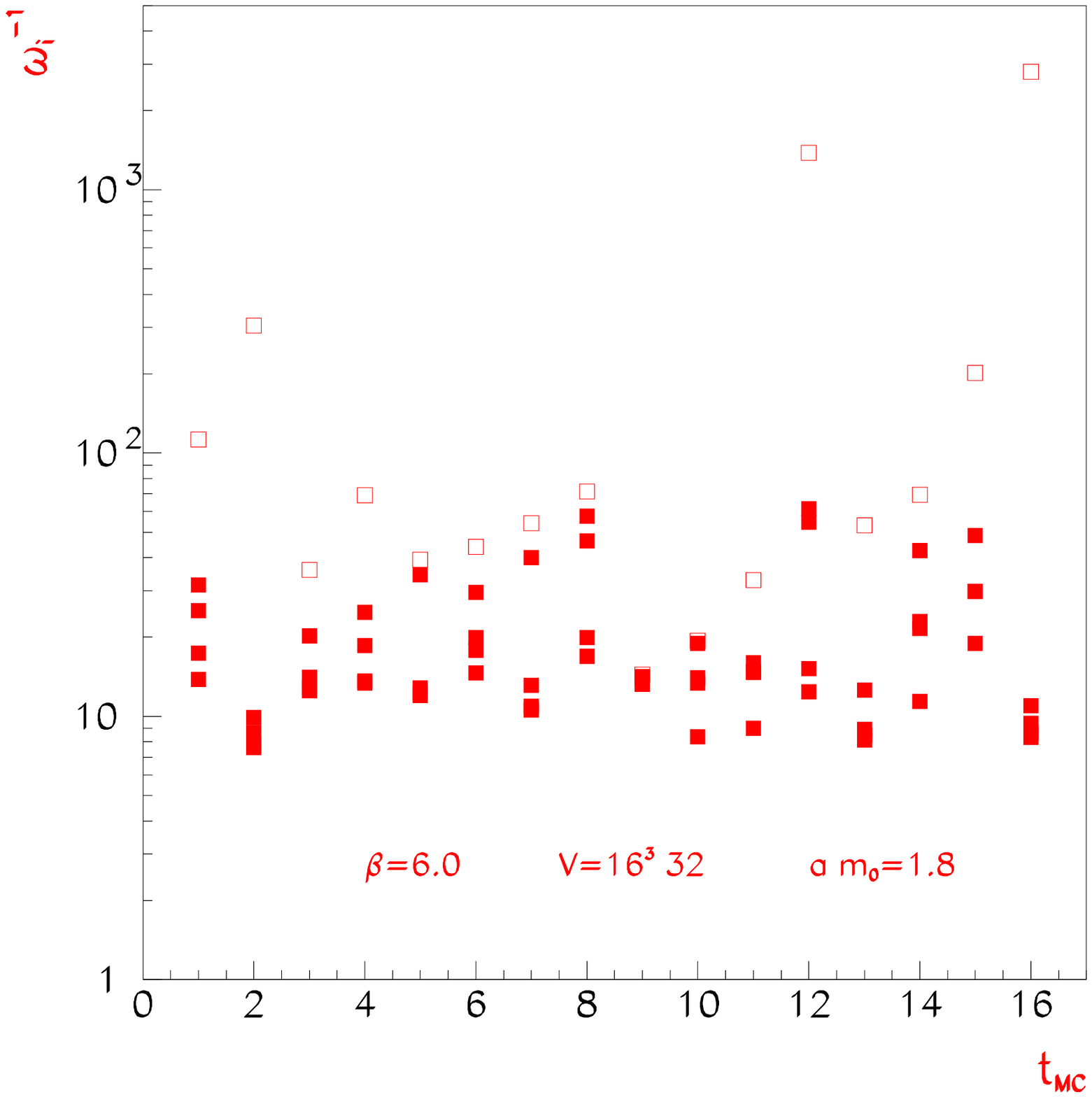, %
width=11cm,height=11cm}
\end{center}
\caption{ \label{fig:eigb60} The same as Fig.~\ref{fig:eigb585} at $\beta=6$ on a $16^3\cdot 32$ lattice.}
\end{figure}

We have studied numerically the convergence rate in the quenched approximation.  We find that it is always controlled by the lowest eigenvalue $\lambda_\mathrm{min}$ of $A^\dagger A$. In Figs.~\ref{fig:eigb585} and \ref{fig:eigb60} 
we show the inverse convergence rates $\omega_i^{-1}$ corresponding to the
five lowest eigenvalues of $A^\dagger A$ at $\beta=5.85$ on an $8^3\cdot 16$ lattice, and at $\beta=6.0$ on a $16^3\cdot 32$ lattice, respectively. In both cases we have set $a_s m_0 = a m_0 = 1.8$, which is a typical value used in 
previous simulations.

%%%  Fig. 1
%%%%%%%%%%%%%%%%%%%%%%%%%%%%%FIGURE%%%%%%%%%%%%%%%%%%%%%%%%%%%%%%%%%%%
%%%%%%%%%%%%%%%%%%%%%%%%%%%%%%%%%%%%%%%%%%%%%%%%%%%%%%%%%%%%%%%%%%%%%%

%%%  Fig. 2
%%%%%%%%%%%%%%%%%%%%%%%%%%%%%FIGURE%%%%%%%%%%%%%%%%%%%%%%%%%%%%%%%%%%%
%%%%%%%%%%%%%%%%%%%%%%%%%%%%%%%%%%%%%%%%%%%%%%%%%%%%%%%%%%%%%%%%%%%%%%

Figures~\ref{fig:eigb585} and \ref{fig:eigb60} give a rather pessimistic view
of the convergence of domain-wall fermions to the exact operator: they imply
that several hundred or even thousand slices in the extra dimension would be
needed to achieve a reasonable approximation. Clearly, this would render domain-wall
fermions impracticable. Our results are consistent with the
findings in \cite{scri}. 

It should be noted, however, that a very similar
situation was found for Neuberger's operator
\cite{hjl}. Also in this case
very small eigenvalues of the corresponding $A^\dagger A = M^\dagger M$ occur, turning the computation of its inverse square root extremely costly.

\vspace{0.8cm}

{\em Acceleration of convergence}

\vspace{0.8cm}

In the case of Neuberger's operator, the bad convergence behaviour
resulting from the low-lying eigenvalues of $M^\dagger M$ could be  cured
by treating these modes exactly
\cite{florida,uns}. 
It is natural to look for  a similar trick also for domain-wall fermions, given the similarity of the two constructions.  
We have found two ways of achieving this. The first method is equivalent
to the one described in \cite{scri}, so we will not give any details here.
The corresponding improved 5D operator differs from the 
standard one by boundary terms. 
%, which we developed independently, while
%the second method follows a different, alternative path.

We have tested the inversion of this
 improved operator, $\mathcal{D}_\mathrm{impr}$, 
by solving the linear equation  
$\mathcal{D}_\mathrm{impr}X=Y$ for a given 
source $Y$. As a numerical solver we have used a conjugate gradient
method with a stopping criterion
$\|r\|/\|X\|<\epsilon$, where $r=\mathcal{D}_\mathrm{impr}X-Y$ is the residual vector 
and $\epsilon$ was set to $\epsilon = 10^{-8}$. We found that
when using such a relatively low value of $\epsilon$ the conjugate gradient method
behaves very poorly: for a number of configurations at $\beta=5.85$ on an $8^3\cdot16$ lattice, 
the norm of the residual vector developed a very long tail at rather small
values of $\|r\| \leq \mathrm{O}(10^{-5})$. 
This resulted in a very large number of iterations
in the conjugate gradient algorithm before it 
converged to the desired accuracy. We suspect that some subtle cancellations 
occur in the improved operator leading to unexpectedly large rounding error effects.

Since this behaviour of the conjugate gradient algorithm was rather unsatisfactory,
we developed an alternative improvement method. 
The key observation for the new improved 5D operator is that the relations
in eq.~(\ref{A}) and eq.~(\ref{D}) hold true for {\em any} choice of $M$ as long as 
\begin{equation}
M^\dagger =\gamma_5M\gamma_5\;,\;\;\;\mathrm{det}(2+a_sM)\ne 0\;.
\label{Mproperties}
\end{equation}
This fact may be used to construct an improved $M$ 
for which the very low eigenvalues of $A^\dagger A$ disappear.
A possible form of $M$ that achieves this is given by
\begin{equation}
a_s\hat{M} = a_sM-\sum_{k,l=1}^{r}X_{kl}w_k\otimes w_l^\dagger\gamma_5,\; 
\label{Mimproved}
\end{equation}
%The corresponding improved 5D operator is evaluated by the following steps.
where 
\begin{equation}
\gamma_5 A v_k=\alpha_kv_k, \;\;\;\;\;k=1,\cdots,r,\;\;\;\;\;(v_k,v_l)=\delta_{kl}\; .
\label{eigenA}
\end{equation}
and 
\begin{equation}
w_k=(2+a_sM)\gamma_5v_k.
\label{defw}
\end{equation}
Finally  
\begin{equation}
(X^{-1})_{kl} = 2\delta_{kl}(\hat{\alpha}_k-\alpha_k)^{-1} + (v_k,(2+a_sM)\gamma_5v_l)\; .
\label{X}
\end{equation}
The corresponding 5D operator $\hat{\cal D}$ is given by eq. (\ref{5doperator}) after substituting $M$ by $\hat M$. Notice that the improved operator differs from 
the original one also in the bulk and not just at the boundary. 

After some algebra it can be shown that
\begin{equation}
\hat{A}\equiv -a_s\hat{M}(2+a_s\hat{M})^{-1} = A + \sum_{k=1}^r(\hat{\alpha}_k-\alpha_k)
             \gamma_5v_k\otimes v_k^\dagger\; .
\label{Ahat}
\end{equation}
It is now easy to see that $\gamma_5 \hat{A}$ has the same eigenvectors as $\gamma_5 A$, but all eigenvalues $\alpha_k,\; k=1,\cdots,r,$ are replaced by $\hat{\alpha}_k$. The limit $N\rightarrow \infty$ of the corresponding improved 4D operator ${\hat D}_N$ is the same as that of the original $D_N$ provided $\mathrm{sign}({\hat \alpha}_k) = \mathrm{sign}(\alpha_k)$. However, the approach to this limit is faster for ${\hat D}_N$ if the lowest eigenvalues of ${\hat A}^\dagger {\hat A}$, ${\hat \lambda}_k \equiv {\hat \alpha}^2_k$,  are larger than those of $A^\dagger A$, i.e.
if $|{\hat \alpha}_k| > |\alpha_k|$.

The concrete choice of $|\hat{\alpha}_k|$
has to be taken with some care to optimize the convergence of the inverter. For example, taking $|\hat{\alpha}_k|=1$ 
led to a bad convergence behaviour of the conjugate gradient algorithm. 
It is our experience that choosing $|\hat{\alpha}_k|$ not much higher than 
$|\alpha_r|$,
$r$ being the index of the largest eigenvalue projected out, see eq.~(\ref{eigenA}),
leads to a normal behaviour of the conjugate gradient algorithm.

\begin{figure}
\vspace{0.0cm}
\begin{center}
\psfig{file=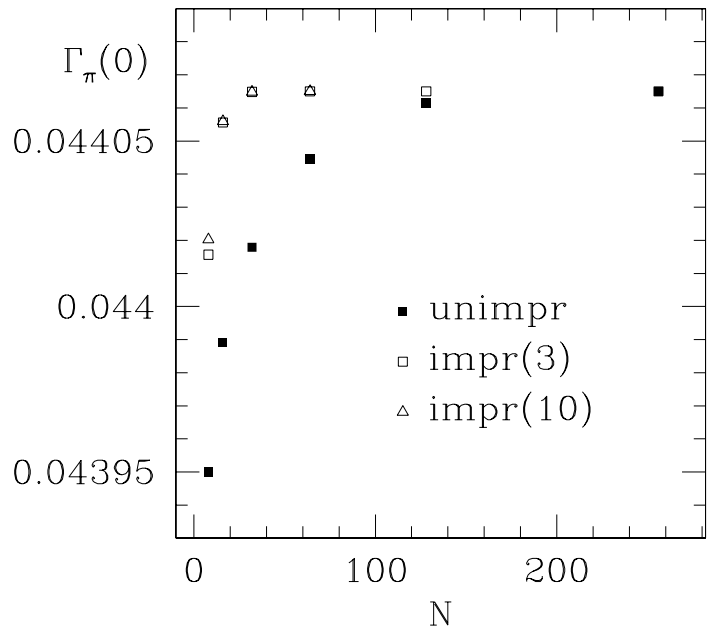, %
width=11cm,height=11cm}
\end{center}
\caption{ \label{fig:pion_0} The pion propagator at zero distance as a function
of $N$. The black squares are the results for the original 5D domain
wall operator. The open symbols correspond to the improved operator with three
and ten eigenvalues projected out. The data are obtained on an $8^3\cdot 16$ lattice at 
$\beta=5.85$ and $am=0.1$.} 
\end{figure}

As an example of the effect of the improvement using $\hat{M}$, eq.~(\ref{Mimproved}),
we show in Fig.~\ref{fig:pion_0} 
the behaviour of the 
pion propagator at zero distance $\Gamma_\pi(0)$  at $\beta=5.85$ on an $8^3\cdot 16$ lattice and for a quark mass, $a m =0.1$.
A similar behaviour is obtained for the pion propagator at other distances. 

%%%  Fig. 1
%%%%%%%%%%%%%%%%%%%%%%%%%%%%%FIGURE%%%%%%%%%%%%%%%%%%%%%%%%%%%%%%%%%%%
%%%%%%%%%%%%%%%%%%%%%%%%%%%%%%%%%%%%%%%%%%%%%%%%%%%%%%%%%%%%%%%%%%%%%%
 
Already the projection of only three low-lying eigenvalues is sufficient to accelerate 
the convergence substantially: similar approximations to the $N\rightarrow \infty$ limit are obtained for $N \sim 150$ in the unimproved case and 
$N\sim 30$ in the improved one. It would, of course,
be interesting to see the effect also on other physical quantities.
 
\vspace{0.8cm}

{\em Conclusions}
 
\vspace{0.8cm}

In this note we presented numerical evidence that in practical simulations
domain wall fermions need an unacceptably large number of slices in the
extra dimension to ensure that the chiral symmetry-breaking terms are 
suppressed. The reason is that 
very small eigenvalues of a 
4D operator appear, which are directly related to the convergence
rate of the 5D domain-wall operator. These results confirm the findings in \cite{scri}.

As in the case of Neuberger's operator, it is possible, however, to 
separate a number of eigenvalues of the 4D operator and treat
them exactly or shift them to larger harmless values. We tested two 
different implementations of this idea. The first has already appeared 
in \cite{scri}, the second, which is described in detail above, is new. 
We found that numerically the second implementation works much better. 
%We propose therefore to use this version of an improved 5D 
%operator instead of the method given in \cite{scri}.

It is our overall impression, however, that there is no particular
advantage in using domain-wall fermions instead of Neuberger's 
operator. Theoretical considerations demonstrate that 
both approaches to realize a chiral symmetry on the lattice are 
equivalent, to the extent that they satisfy the Ginsparg--Wilson relation. 
However, in practical applications,  it is our present experience that 
it is 
easier to control chiral symmetry violations with Neuberger's operator.
 
\vspace{0.8cm}

{\em Acknowledgements}\\
We would like to thank Giulia de Divitiis for her help at an early stage
of this project and the computer centres at NIC 
(J\"ulich) and CIEMAT (Madrid) for providing computer time and technical
support.


\begin{thebibliography}{99}
%
\bibitem{kaplan} D.B.~Kaplan, Phys. Lett. {\bf B288} (1992) 342.
%
\bibitem{blum0} T.~Blum, Nucl. Phys. (Proc.Suppl.) {\bf B73} (1999) 167.
%
%\bibitem{nieder} F. Niedermayer, Nucl. Phys. (Proc.Suppl.) {\bf B73} (1999) 105.
%
\bibitem{shamir1} Y.~Shamir, Nucl. Phys. {\bf B406} (1993) 90.
%
\bibitem{shamir2} V.~Furman and Y.~Shamir, Nucl. Phys. {\bf B439} (1995) 54.
%
\bibitem{neub3} H.~Neuberger, Phys. Rev. {\bf D57} (1998) 5417.
%
%  In this paper the determinant of the dw is worked out and the relation to 
% the determinant of the 4D operator derived
%
\bibitem{kiku1} Y.~Kikukawa and T.~Noguchi, hep-lat/9902022.
%
\bibitem{gw} P.H.~Ginsparg and K.G.~Wilson, Phys. Rev. {\bf D25} (1982) 2649.
%
\bibitem{hasen2} P.~Hasenfratz, Nucl. Phys. B (Proc.Suppl.) {\bf 63A-C} (1998) 53.
%
\bibitem{neub2} H.~Neuberger, Phys. Lett. {\bf B427} (1998) 353.
%
% Here this GW relation for Neuberger's operator is shown
%
\bibitem{luscher} M.~L\"uscher, Phys. Lett. {\bf B428} (1998) 342.
%
\bibitem{hasen} P.~Hasenfratz, V.~Laliena and F.~Niedermayer, Phys. Lett. {\bf B427} (1998) 125.
%
\bibitem{hasen3} P.~Hasenfratz, Nucl. Phys. {\bf B525} (1998) 401. 
%
\bibitem{chandra} S.~Chandrasekharan, Phys. Rev. {\bf D60} (1999) 074503.
%
\bibitem{neub1} H.~Neuberger, Phys. Lett. {\bf B417} (1998) 141.
%
% Here the famous operator is writen
% 
\bibitem{shamir3} Y.~Shamir, hep-lat/0003024.                 
%
\bibitem{blum1} T.~Blum, A.~Soni and M.~Wingate, Nucl. Phys. B (Proc.Suppl.) {\bf 73} (1999) 201.
%
\bibitem{blum2} T.~Blum and A.~Soni, Phys. Rev. Lett. {\bf 79} (1997) 3595.
%
\bibitem{cppacs} A.~Ali Khan, {\it et al.} (CP-PACS Collab.), hep-lat/9909049.
%
\bibitem{scri} R.G.~Edwards and U.~Heller, hep-lat/0005002.
%
\bibitem{borici} A.~Borici, Phys. Lett. {\bf B453} (1999) 46.
%
\bibitem{kiku2} Y.~Kikukawa, hep-lat/9912056.
%
\bibitem{hjl} 
P.~Hern\'andez, K.~Jansen and M.~L\"uscher, Nucl. Phys. {\bf B552} (1999) 363.  
%
\bibitem{martin} M.~L\"uscher, Nucl. Phys. {\bf B538} (1999) 515.
%
%
\bibitem{cg}
B.~Bunk, K.~Jansen, M.~L\"uscher and H.~Simma, DESY report (September 1994);
T.~Kalkreuter and H.~Simma, Comput. Phys. Commun. {\bf 93} (1996) 33.
%
\bibitem{florida} R.G.~Edwards, U.M.~Heller and R.~Narayanan, Phys. Rev. {\bf D59} (1999) 094510.
%
\bibitem{uns} 
       P.~Hern\'andez, K.~Jansen and L.~Lellouch, 
       Phys. Lett. {\bf B469}(1999) 198; hep-lat/0001008.
%

%
%\bibitem{numrec}
%W.H.~Press, S.A.~Teukolsky, W.T.~Vetterling and B.P.~Flannery,
%{\em Numerical Recipes},
%Second Edition, Cambridge University Press, Cambridge, 1992.
%
%
%\bibitem{florida2} R.G.~Edwards {\it et al.}, hep-lat/9902117.

\end{thebibliography}
\end{document}